\documentclass[a4paper,fleqn,usenatbib]{mnras}
\usepackage{amssymb,amsmath,graphicx,psfrag,mathtools}
\usepackage[T1]{fontenc} %MNRAS
\usepackage{ae,aecompl} %MNRAS
\usepackage{url}
\usepackage{revsymb}
\hypersetup{draft} %TO AVOID ERROR: \PDFENDLINK ENDED UP IN DIFFERENT
\title[Precession and Jitter in FRB 180916B]{Precession and Jitter in FRB
180916B}
\author[J. I. Katz]{
J. I. Katz,$^{1}$\thanks{E-mail katz@wuphys.wustl.edu} %MNRAS
\\
$^{1}$Department of Physics and McDonnell Center for the Space Sciences,
Washington University, St. Louis, Mo. 63130 USA %MNRAS
}
\date{Accepted XXX.  Received YYY; in original form ZZZ} %MNRAS
\pubyear{2022} %MNRAS
\date{\today}
\begin{document} %MNRAS
%\psfrag{theta}{$\theta$}  Works only on COMPLETE strings
\label{firstpage} %MNRAS
\pagerange{\pageref{firstpage}--\pageref{lastpage}} %MNRAS
\maketitle %MNRAS
\begin{abstract}
Recent CHIME/FRB observations of the periodic repeating FRB 180916B have
produced a homogeneous sample of 44 bursts.  These permit a redetermination
of the modulation period and phase window, in agreement with earlier
results.  If the periodicity results from the precession of an accretion
disc, in analogy with those of Her X-1, SS 433, and many other superorbital
periods, the width of the observable phase window indicates that the disc
axis jitters by an angle of about 0.14 of the inclination angle, similar to
the ratio of 0.14 in the well-observed jittering jet source SS 433.
\end{abstract}
 %MNRAS
\section{Introduction}
The bursts of the repeating FRB 180916B have been shown to be confined to
a window about five days wide that repeats with a period of 16.3 days
\citep{CHIME20,P21}.  A period of 160 days has also been suggested for FRB
121102 \citep{R20}.  A number of models of this behavior have been proposed,
but none have yet been definitively tested \citep{K21a}; \citet{Mck22}
provide references to the original literature.

The recent extended study of FRB 180916B by \citet{Mck22} reported 44 bursts
observed by CHIME/FRB.  This sample is expected to be homogeneous because
CHIME/FRB is an unsteered transit instrument whose observations do not
depend on a choice of observing times thought to be propitious for
detection, or other times for other reasons.  There are several likely
contributions to variations in sensitivity: varying amplifier gain, a
varying number of feeds down, varying weather, varying electromagnetic
interference, changes in detection threshold and shutdowns for maintenance
or upgrades \citep{Mckp}.  These are unlikely to be correlated with phase in
the 16 day cycle and therefore do not affect the mean dependence of burst
activity on phase, although they add stochastic noise by reducing the number
of observed bursts.  Were they correlated with 16 day phase, we would not
accept the reality of the 16 day period.

This paper investigates the implications of these data for the hypothesis
\citep{K20,S21} that the periodicity results from the emission of bursts
along the rotation axis of a precessing accretion disc in a binary system.
Precessing discs are observed in Her X-1/HZ Her, the prototype (type
specimen or holotype in biological nomeclature) superorbital binary with an
orbital period of 1.7 d and a superorbital period of 35 d \citep{K73} in
which X-rays are emitted by a neutron star, in SS 433 with an orbital
period of 5.6 d and a superorbital period of 164 d \citep{K80} in which an
accretion disc around a black hole launches a sub-relativistic jet along its
axis, and in many other mass-transfer binaries.  Attributing periodicity to
disc precession follows from the hypothesis \citep{K17} that repeating FRB
are produced in accretion disc funnels.

Fast radio bursts (FRB) have not been detected from known superorbital
binaries, but the line of sight is aligned with the disc axis in only a
small fraction of such objects, and only intermittently because the disc
axis is, in at least some of them, precessing and jittering, so this might
be an observational selection effect.  Alternatively, those emitting FRB may
be distinguished from those that do not in some other, as yet not
understood, manner \citep{K21b}: obvious candidates include the mass and
nature (neutron star or black hole) of the central object, accretion rate,
turbulence in the accretion flow, and the magnetic field in the accreting
matter.

{The periodicity and period of FRB 180916B are well established
\citep{CHIME20,P21,Mck22}.  The chief purposes of this study were to examine
the distribution of burst phases about exact periodicity, to compare to
observations of SS 433, and to investigate the applicability of the
jittering precessing beam model to FRB 180916B.}
\section{The Period}
The times of the 44 bursts observed by \citet{Mck22} may be used to
redetermine the modulation period.  One way of doing this is to calculate
the periodogram of the burst times.  This is defined by
\begin{equation}
        A(P) = \sqrt{C^2(P)+S^2(P)},
\end{equation}
where
\begin{equation}
        \begin{split}
                C(P) &= \sum_n \cos{2 \pi T_n/P}\\
                S(P) &= \sum_n \sin{2 \pi T_n/P},
        \end{split}
\end{equation}
where $T_n$ is the time of the $n$-th burst and $P$ is the period.

\begin{figure}
	\centering
	\includegraphics[width=3in]{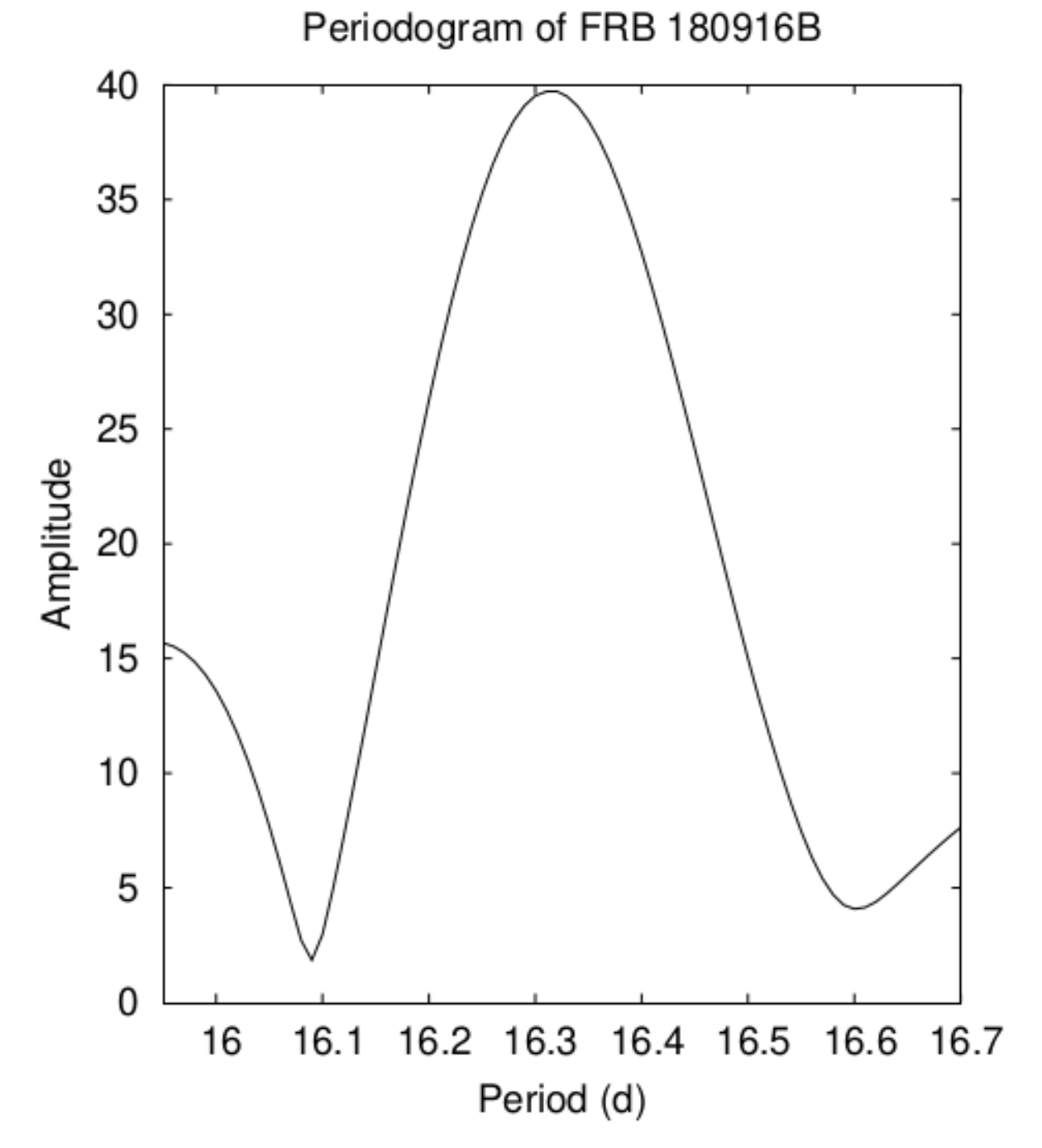}
	\caption{\label{periodogram}Periodogram of the 44 bursts of FRB
	180916B observed by \citet{Mck22}.}
\end{figure}

The periodogram of these 44 bursts {for $15.95\,\text{d} \le P \le
16.71\,\text{d}$} is shown in Fig.~\ref{periodogram}.
The best estimate of the period is the value for which the periodogram is
maximum, $P = 16.315\,$d.  This is close to and consistent with values
obtained from previous data by \citet{CHIME20,P21}.
\section{Burst Phases}
\begin{figure}
	\centering
	\includegraphics[width=3in]{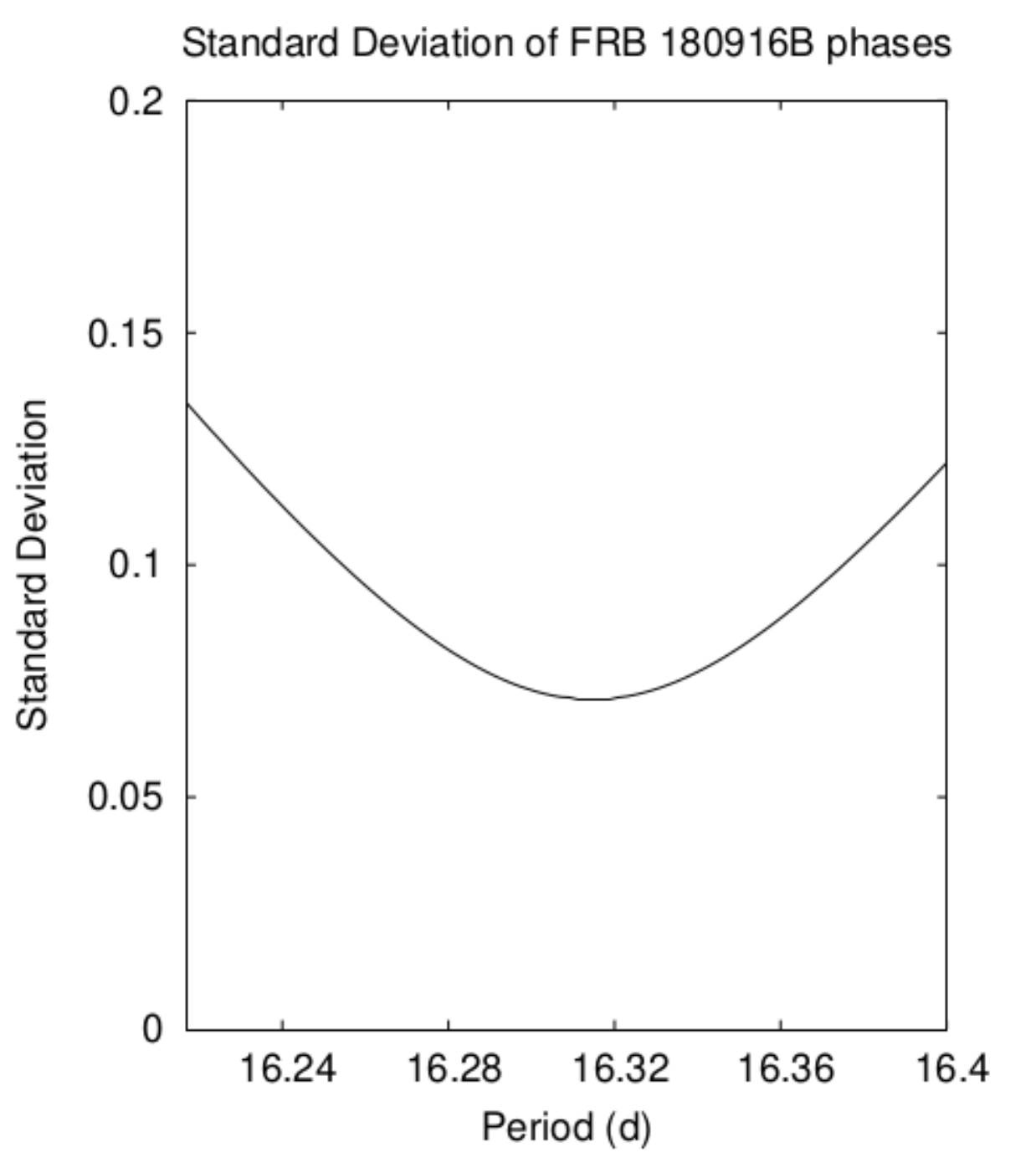}
	\caption{\label{variance} Standard deviation of burst phases from
	constant $P$ ephemeris, as a function of $P$.}
\end{figure}

Alternatively, the mean and the standard deviation of the phases of the
individual bursts may computed as functions of the period.  The results are
shown in Fig.~\ref{variance}.  Unsurprisingly, the standard deviation is a
minimum for the same $P = 16.315\,$d as the maximum of the periodogram.

\begin{figure}
	\centering
	\includegraphics[width=3in]{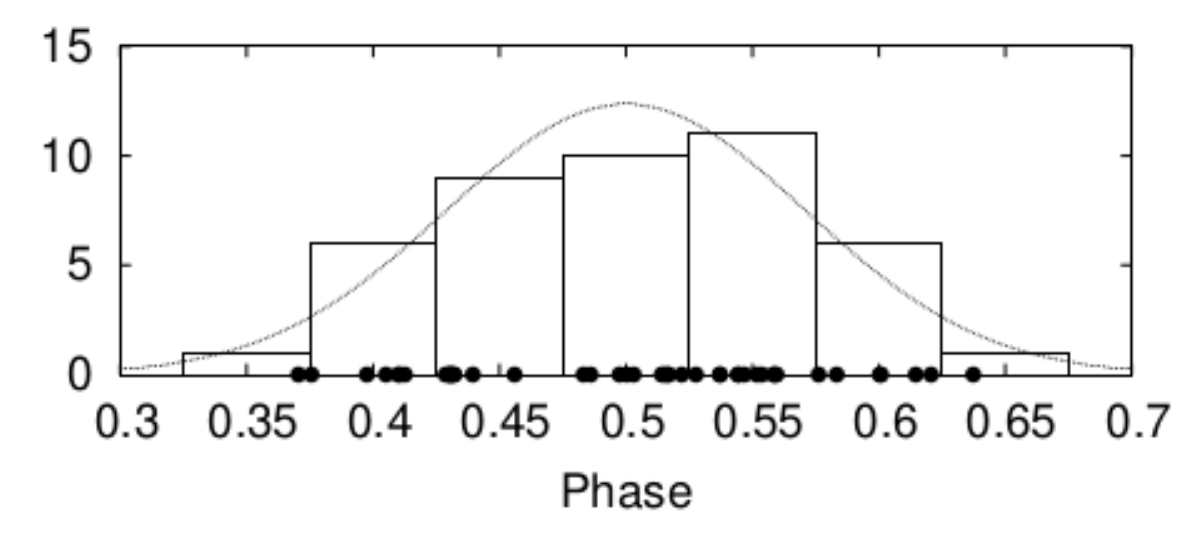}
	\caption{\label{phases} Phases of the 44 individual bursts observed
	by \citet{Mck22} for the fitted period of 16.315 d (points), a bar
	graph of their distribution and (dotted) the fitted Gaussian.  The
	mean phase is defined as 0.5 and their standard deviation is 0.071.}
\end{figure}

The phases of the 44 bursts, with phase 0.5 defined as the mean phase with
the best-fit period, are shown in Fig.~\ref{phases}.  Their standard
deviation is 0.071 cycles, or 1.16 d.  The period, assuming a Gaussian
random distribution of phase offsets, is ($\pm 1\sigma$ error estimate)
\begin{equation}
	P = 16.315 \pm 0.175\ \text{d},
\end{equation}
consistent with earlier determinations \citep{CHIME20,P21}.  The observed
distribution is consistent with the Gaussian fit.

\section{Jitter}
One possible model of the periodicity of FRB 180916B attributes it to
emission along the angular momentum axis of an accretion disc that
precesses as a result of the torque exerted by a mass in orbit around the
accreting object.  The most quantitatively-studied such system is the binary
SS 433.  The masses of its components have long been uncertain, but the most
recent determination \citep{B18} indicates the accreting object is a black
hole of mass $15 \pm 2\,M_\odot$, and its stellar companion has a mass of
about $20\,M_\odot$.

The axis of the precessing disc of SS 433 jitters around its mean precession
\citep{KP82,I93,CG94,Ku10}.  It is uniquely well observed because its
sub-relativistic jet of ionized gas, assumed to be emitted along its axis,
radiates emission lines whose Doppler shifts are measured quantitatively
\citep{M81,M84,F04}.  Observations of other precessing discs \citep{LJ82}
also suggest jitter about their mean motion.

{We assume the FRB, and the jet in SS 433, are emitted along the
instantaneous angular momentum axis of their discs and that any deviation be
small compared to the amplitude of the jitter.}
The distribution of jitter angles in SS 433 is obtained directly from the
jet's Doppler shifts \citep{KP82,I93,CG94,Ku10}.  Its half-width
$\Delta\theta \approx 0.05\,$radian.  The precession angle $\theta \approx
20^\circ \approx 0.35\,$radian and $\Delta\theta/\theta \approx 0.14$.  In
the proposed model of FRB 180916B the observed standard deviation of phase
$\Delta\phi \approx 2 \pi \times 0.071\,$radian implies (because 0.071 is
the half-width in phase) $\Delta\theta/\theta \approx (\Delta\phi/\pi)
\approx 0.14$.  The equality to the corresponding value for SS 433 is surely
fortuitous, but the fact that they are comparable is consistent with the
proposed model.
\section{Discussion}
The observation \citep{PM21} that the activity window is wider at low
frequencies is consistent with emission in an accretion funnel.  Lower
frequency waves may be refracted away from the funnel axis by the plasma of
the jet, and lower frequency emission may be produced further from the
central compact object where the angular width of the funnel is wider
\citep{S21}.

The known object most closely resembling the model presented here is SS 433;
we are always far from its beam axis \citep{M81,M84,F04} so the absence of
FRB from SS 433 does not disprove the model of FRB emission along a disc
axis.  Nor does the identification of FRB 200428 with a Soft Gamma Repeater
that does not appear to have an accretion disc, if repeating and apparently
non-repeating FRB are produced by different classes of sources \citep{K22}.
At extra-Galactic distance FRB 200428, if close enough to be observed at
all, would be an apparent non-repeater, unlike the repeating FRB 180916B.
The hypothesis that repeating FRB are emitted along a disc axis is
independent of the nature of the object emitting the FRB, and does not
require that the disc precess; precession is only required when FRB activity
is modulated periodically, as it is in FRB 180916B.

The rarity of objects like SS 433 (there appears to be only one in our
Galaxy), combined with the fortuitous alignment required to observe emission
collimated with the axis of a precessing jet, may explain the low density of
{\it observed\/} repeating FRB in the Universe.  This is difficult to
quantify because of the uncertain abundance of such possible FRB sources and
the uncertain accuracy of alignment and duration and sensitivity of
observation required for their discovery.

The binary precessing disc model is the only model of the periodic
modulation of FRB activity that has been observed in another object.  The
results presented here constrain its parameters if it is correct.
\section*{Acknowledgements}
I thank R. Mckinven and T. Piran for useful discussions.
\section*{Data Availability}
This theoretical study did not generate any new data.

\label{lastpage} %MNRAS

\begin{thebibliography}{99}
	\bibitem[\protect\citeauthoryear{Bowler}{2018}]{B18} Bowler, M. G.
		2018 \aap\ 619, L4.
	\bibitem[\protect\citeauthoryear{CHIME/FRB Collaboration}{2020}]
		{CHIME20} CHIME/FRB Collaboration 2020 Nature 582, 351.
	\bibitem[\protect\citeauthoryear{Collins \& Garasi}{1994}]{CG94}
		Collins, G. W. II \& Garasi, C. J. 1994 \apj\ 431, 836.
	\bibitem[\protect\citeauthoryear{Fabrika}{2004}]{F04} Fabrika, S.
		2004 Astrophysics Space Physics Rev. 12, 1.
	\bibitem[\protect\citeauthoryear{Iijima}{1993}]{I93} Iijima, T.
		1993 \apj\ 410, 295.
	\bibitem[\protect\citeauthoryear{Katz}{1973}]{K73} Katz, J. I.
		1973 Nature Phys. Sci. 246, 87.
	\bibitem[\protect\citeauthoryear{Katz}{1980}]{K80} Katz, J. I.
		1980 \apjl\ 236, L127.
	\bibitem[\protect\citeauthoryear{Katz}{2017}]{K17} Katz, J. I.
		2017 \mnras\ 471, L92.
	\bibitem[\protect\citeauthoryear{Katz}{2020}]{K20} Katz, J. I.
		2020 \mnras\ 494, L64.
	\bibitem[\protect\citeauthoryear{Katz}{2021a}]{K21a} Katz, J. I.
		2021a \mnras\ 502, 4664.
	\bibitem[\protect\citeauthoryear{Katz}{2021b}]{K21b} Katz, J. I.
		2021b \mnras\ 508, L12.
	\bibitem[\protect\citeauthoryear{Katz}{2022}]{K22} Katz, J. I.
		2022 arXiv:2203.03675.
	\bibitem[\protect\citeauthoryear{Katz \& Piran}{1982}]{KP82}
		Katz, J. I. \& Piran, T. 1982 Ap.~Lett. 23, 11.
	\bibitem[\protect\citeauthoryear{Kubota {\it et al.\/}}{2010}]
		{Ku10} Kubota, K., Ueda, Y., Kawai, N. {\it et al.\/} 2010
		\pasj\ 62, 323.
	\bibitem[\protect\citeauthoryear{Levine \& Jernigan}{1982}]{LJ82}
		Levine, A. M. \& Jernigan, J. G. 1982 \apj\ 262, 294.
	\bibitem[\protect\citeauthoryear{Margon}{1984}]{M84} Margon, B.
		1984 \araa\ 22, 507.
	\bibitem[\protect\citeauthoryear{Mckinven}{2022}]{Mckp} Mckinven, R.
		2022 personal communication.
	\bibitem[\protect\citeauthoryear{Mckinven {\it et al.\/}}{2022}]
		{Mck22} Mckinven, R., Gaensler, B. M., Michilli, D. {\it et
		al.\/} 2022 arXiv:2205.09221.
	\bibitem[\protect\citeauthoryear{Milgrom}{1981}]{M81} Milgrom, M.
		1981 Vistas Astr. 25, 141
	\bibitem[\protect\citeauthoryear{Pastor-Marazuela {\it et al.\/}}
		{2021}]{PM21} Pastor-Marazuela, I., Connor, I., van Leeuwen,
		J. {\it et al.\/} 2021 Nature 596, 505.
	\bibitem[\protect\citeauthoryear{Pleunis {\it et al.\/}}{2021}]{P21}
		Pleunis, Z., Michilli, D., Bassa, C. G. {\it et al.\/} 2021
		\apjl\ 911, L3.
	\bibitem[\protect\citeauthoryear{Rajwade {\it et al.\/}}{2020}]{R20}
		Rajwade, K. M., Mickaliger, M. B., Stappers, B. W. {\it et
		al.\/} 2020 \mnras\ 495, 3551.
	\bibitem[\protect\citeauthoryear{Sridhar {\it et al.\/}}{2021}]{S21}
		Sridhar, N., Metzger, B. D., Beniamini, P. {\it et al.\/}
		2021 \apj\ 917, 13.
\end{thebibliography}
\end{document}